\documentclass[pre,twocolumn,showpacs,preprintnumbers,amsmath,amssymb]{revtex4}

\usepackage{epsfig,graphicx}
\usepackage{dcolumn}
\usepackage{bm}
\usepackage{amsmath}
\usepackage{times}
\usepackage{mathptm}

\def\s#1{_{\rm #1} }
\def\sp#1{^{\rm #1} }
\def\Tr#1{{\rm Tr}\left( #1 \right)}

\def\deltam{\matr{\delta}}

\def\n{\ensuremath{{\bf{n}}}}
\def \be{\begin{equation}}
\def \bea{\begin{eqnarray}}
\def \ee{\end{equation}}
\def \eea{\end{eqnarray}}
\def\vep{\varepsilon}
\def\vec#1{{\rm\bf #1}}

\def\R{{\vec{R}}}

\def\x{\hat{\vec{x}} } 
\def\y{\hat{\vec{y}}}
\def\z{\hat{\vec{z}}}
\def\k{\vec{k} }
\def\m{\vec{m} }

\def\matr#1{\ensuremath{\underline{\underline{{{#1}}}}}}
\def\lz{\ell_{\parallel}}
\def\lp{\ell_{\bot}}
\def\lm{ \matr{\lambda} }
\def\llm{ \matr{\ell} }
\def\half{\textstyle \frac{1}{2}}

\def\Free{\half \mu \Tr{\llm\s{0}\cdot \lm^{\rm T}\cdot \llm^{-1}\cdot \lm }}

\def\lambdam{ \lambda_{\rm m} }

\begin{document}

\title{Uniaxial and biaxial soft deformations of nematic elastomers}
\author{M. Warner and S. Kutter}
 \affiliation{Cavendish
Laboratory, University of Cambridge, Madingley Road, Cambridge CB3
0HE, U.K.} \date{\today}


\begin{abstract} 
We give a geometric interpretation of the soft elastic deformation 
modes of nematic elastomers, with explicit examples, for both 
uniaxial and biaxial nematic order.  We show the importance of 
body rotations in this non-classical elasticity and how the 
invariance under rotations of the reference and target states 
gives soft elasticity (the Golubovic and Lubensky theorem).  The 
role of rotations makes the Polar Decomposition Theorem vital for
decomposing general deformations into body 
rotations and symmetric strains.  The role of the square 
roots of tensors is discussed in this context and that of finding 
explicit forms for soft deformations (the approach of Olmsted). 
\end{abstract}


  
\maketitle

\section{Introduction}

Nematic elastomers display three unique and related phenomena not 
found in conventional elasticity - large spontaneous deformations, 
very large optical-mechanical response and soft elasticity.  The 
last, shape change with little or no energy cost, is the subject 
of this paper.  We give a geometric interpretation of the soft 
modes described using fractional powers of tensors.  We then 
discuss the soft modes of biaxial nematic elastomers.  Since in 
contrast to conventional elastic solids, rotations play an 
essential role in the elasticity of nematic rubber, we conclude by 
discussing the related questions of breaking finite strains into 
symmetric shears and rotations, the Polar Decomposition Theorem 
for tensors, and the nature of square roots of tensors.

Nematic elastomers have an internal, orientational degree of 
freedom  in addition to those of ordinary elastic bodies.  The 
anisotropy of molecular orientation induces shape anisotropy in 
the polymers that make up the elastomer.  The switching on and off 
of this molecular shape anisotropy, either by temperature 
change\cite{Tshape} or by illumination\cite{photoelastomers_1}, 
causes large ($\approx 400\%$) mechanical shape changes. 

Rotation of this anisotropy, by imposed mechanical strains, is the 
extreme opto-mechanical effect that is observed.  When  
accompanied by subsidiary shears and contractions which 
accommodate the changes of molecular distributions, the cost of 
the originally imposed strain is rendered to zero .  This is what 
we call soft elasticity\cite{WarnerJPhysII}. 

\section{Softness in linear continua}
One can explore the related ideas of anisotropy, rotation and soft 
elasticity initially for small deformations and rotations, that is 
within the linear continuum theory of a uniaxial body with a 
mobile director $\n$ that characterises the anisotropy 
direction\cite{deG_1981}. The free energy density, $F$, is: 
\begin{eqnarray}
    F &=&  \half\tilde{B} ( {\rm Tr} [\matr{ \tilde{ \varepsilon } } ]   
        )^2 + \half \kappa \ {\rm Tr} [\matr{ \tilde{ \varepsilon } } 
        ]({\bf n}\cdot \matr{\vep}\cdot {\bf n}) \ + \half \mu_0 ({\bf 
        n}\cdot \matr{\vep}\cdot {\bf n})^2 \nonumber\\
    &&+ \half \mu_1 
        [{\bf n}\times \matr{\vep} \times {\bf n}]^2 + \half \mu_2 ([{\bf 
        n}\times \matr{\vep}] \cdot {\bf n})^2 \label{peterF} \\
    &&+ \half D_1 [(\bm{\Omega}-\bm{\omega})\times {\bf 
        n}]^2 + \half D_2 \, {\bf n}\cdot \matr{\vep} \cdot [ 
        (\bm{\Omega}-\bm{\omega})\times {\bf n}],  \nonumber
\end{eqnarray}
where $\matr{\varepsilon}=\matr{\tilde{\varepsilon}}-
\frac{1}{3}{\rm Tr}[\matr{\tilde{\varepsilon}}]\ \matr{\delta}$ is
the traceless part of the linear symmetric strain
$\tilde{\varepsilon}_{ij}=\frac{1}{2}(\partial_j u_i+\partial_i u_j)$, and
${\bf \Omega}=\frac{1}{2}{\rm curl}\ {\bf u}$ is the antisymmetric part (the
body rotation), ${\bf u}$ being the displacement field.

\begin{figure}[h]
\centering 
\resizebox{0.45\textwidth}{!}{\includegraphics{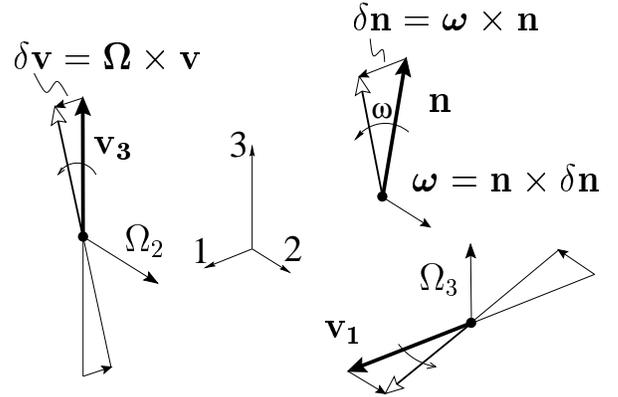}} 
\caption[]{Rotation of $\n$ about an axis $\bm{\omega}$, and body 
rotations $\bm{\Omega}$ rotating vectors $\vec{v}$. Only 
$\Omega_2$ and $\Omega_1$ (not shown) have any effect on vectors 
parallel to $\n$. } \label{directorrot} 
\end{figure}

The latter terms are those of the relative rotation
$(\bm{\Omega}-\bm{\omega})\times {\bf n}$ coupling: for small 
rotations, the  director variation corresponds to a rotation 
$\bm{\omega} = 
{\bf n}\times \delta {\bf n}_{\bm \omega}$ or
$\delta \vec{n}_{\bm\omega} = \bm{\omega}\times \vec{n} $, see 
fig.~\ref{directorrot}.  Similarly, a small rotation $\bm{\Omega}$ of the 
matrix causes a vector $\vec{v}$ in the body to suffer the change 
$\delta \vec{v}_{\bm \Omega} = \bm{\Omega}\times \vec{v}$. 
The net rotation of $\n$ with respect to the matrix inflicted by 
the relative rotation $\bm{\Omega} - \bm{\omega}$ of the matrix 
($\bm\Omega$) relative to the changing director ($\bm\omega$)  
accordingly gives a relative change in  $\n$ of:
\bea 
 \delta \vec{n} =
(\bm{\Omega}-\bm{\omega})\times \n \ .\nonumber
\eea 
The directors $\n$ and $\n_0$  
before and after application of strain are not distinguished in the 
first terms of equation (\ref{peterF}) for small director rotation, 
but clearly must be in the relative rotation coupling terms in 
$F$. 

\begin{figure}[ht] \centering 
\resizebox{0.4\textwidth}{!}{\includegraphics{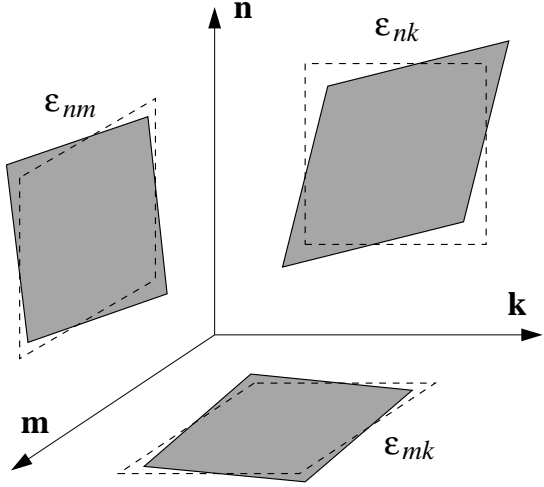}} 
\caption[]{The elements of strain in a uniaxially anisotropic 
medium. Dashed lines show the undistorted state, whereas the strained elements
are shown by shaded areas.
They divide into stretches ${\vep}_{nn}$ along $\n$, stretches 
${\vep}_{mm}$ and ${\vep}_{kk}$ and distortions ${\vep}_{mk}$
in the plane perpendicular to $\n$, and distortions ${\vep}_{nm}$ 
and ${\vep}_{nk}$ encompassing $\n$ and the perpendicular plane.} 
\label{directions} 
\end{figure}

The tensor $\matr{\tilde{\vep}}$ is the strain before it has been 
made traceless, that is, it still has volume changes 
$\Tr{\matr{\tilde{\vep}}}$ in it. Since rubber is a soft material, its 
deformations are at constant volume and we neglect the first two 
terms in equation (\ref{peterF}) since they involve volume change. The 
geometry of the remaining strains is vital and shown in 
fig.~\ref{directions}. 

\begin{figure}[h]
\centering 
\resizebox{0.45\textwidth}{!}{\includegraphics{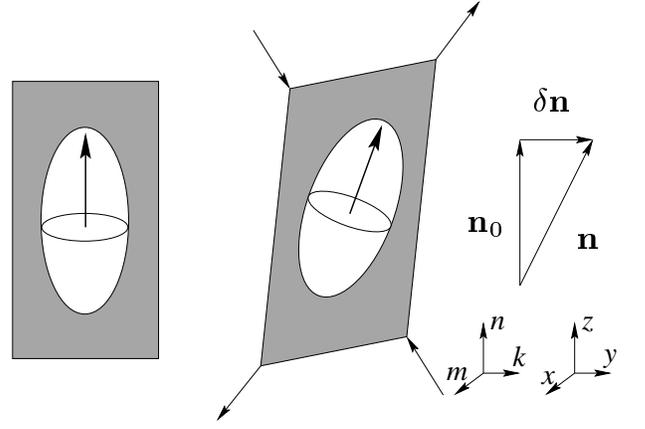}} 
\caption[]{Symmetric shear induces director rotation toward the 
elongation diagonal. The respective elongations and compressions 
along the diagonals are shown. The chain shape distribution also 
rotates. When mechanical shape changes of a network accommodate 
the rotations of the chain distribution without distortions, such 
shape changes can take place at minimal energy cost.} \label{deG2} 
\end{figure}

The de Gennes \cite{deG_1981} relative rotation couplings are 
unique to nematic networks because they require an independent, 
rotational degree of freedom, i.e., the rotations $\bm\omega$ of the director
{\bf n}. Its motion in the medium is coupled to the body rotation 
${\bm\Omega}$. Hence, the elastic energy unusually
involves {\em antisymmetric} components of shear strain.
In section \ref{rss}, we accordingly show how to extract the rotational
component from any finite shear.
The first coupling, 
$D_1$, purely resists the rotation of the director relative to a 
rotating, undeformed matrix. The $D_2$ term couples $\n$ to the 
symmetric part of shear in the plane that involves $\n$ (e.g. 
$\vep_{zx}$ if $n_0=n_z$ in equilibrium).  These are the shears 
${\vep}_{nm}$ and ${\vep}_{nk}$ in fig.~\ref{directions}.  Since 
infinitesimal, incompressible symmetric shear is equivalent to 
stretch along one diagonal and compression along another, it is 
reasonable that prolate molecules will reduce the cost of 
distortion by rotating their ordering direction to being as much 
as possible along the elongation diagonal, depicted in 
fig.~\ref{deG2}.  Oblate elastomers rotate their director toward 
the compression diagonal to achieve the appropriate elastic 
accommodation.  The spheroid represents the anisotropic shape 
distribution of the crosslinked polymers from which the network is 
composed.  

A theorem of Golubovic and Lubensky (GL) \cite{Lubensky1} shows that on 
symmetry grounds any solid with an internal degree of freedom 
which is also capable of reaching an isotropic reference state 
must be invariant under the double set of rotations of both the 
reference and target states when considering elastic deformations.  
This has the remarkable consequence that some elastic deformations 
must be soft.  This was discovered independently and in a 
superficially different form when studying at finite deformations 
the elastic response of nematic elastomers \cite{WarnerJPhysII}.  
We sketch how this can be understood for elastomers in order to 
explain the complex soft modes we shall later discuss. 

An effective shear modulus arises for an imposed elastic strain, 
when the nematic director is free to evolve optimally.  The 
elastic modulus $\mu_2$ is reduced; by the symmetry argument it 
must vanish to give no overall energy cost. If the local rotational
component $\bm{\Omega}$
of the deformation of the elastic matrix  and the rotation of the 
director are coaxial, i.e. $\bm{\Omega}$ and $\bm{\omega}$ are 
parallel and uniform, the argument is simple. Minimising the 
relative-rotation part of the energy density (\ref{peterF}), 
$\frac{1}{2}D_1 [(\bm{\Omega}-\bm{\omega})\times {\bf n}]^2 + 
\frac{1}{2}D_2 \, {\bf n}\cdot {\matr{\vep}} \cdot
[(\bm{\Omega}-\bm{\omega})\times {\bf n}]$, one obtains the optimal 
relative rotation for a given shear strain $\matr{{\varepsilon}}$: 
\begin{equation}
[(\bm{\Omega}-\bm{\omega})\times {\bf n}]= - \frac{D_2}{2 D_1} 
({\bf n} \cdot\matr{{\varepsilon}} )  \ {\rm or} \ \ 
(\bm{\Omega}-\bm{\omega}) =- \frac{D_2}{2 D_1}  [{\bf n} \cdot 
\matr{{\varepsilon}} \times {\bf n}] \ . \label{Oeq} 
\end{equation}
We now substitute this back into the energy density and obtain for 
the rotation-strain terms: 
\begin{eqnarray}
&&\half D_1 [(\bm{\Omega}-\bm{\omega})\times {\bf n}]^2 + \half  
D_2 \, {\bf n}\cdot {\matr{\vep}} \cdot [ 
(\bm{\Omega}-\bm{\omega})\times {\bf n}]  \label{relroteq}\\
&& = - \frac{D_2^2}{8 D_1} [{\bf n}
\cdot \matr{{\varepsilon}} \times {\bf n}]^2 \equiv - 
\frac{D_2^2}{8 D_1} \left( {{\vep}_{xz}}^2 +{{\vep}_{yz}}^2  
\right) \; . \nonumber 
\end{eqnarray}
The last expression is written in the specific coordinate frame 
where the initial director ${\bf n}_0$ is parallel to $z$-axis. We 
can now unite this expression with the rest of the elastic energy 
density, equation (\ref{peterF}) and thus obtain the effective 
rubber-elastic energy which depends only on strains; the director 
does not appear.  For instance, in the specific coordinates of 
fig.~\ref{directions}: 
\begin{eqnarray}
F &=&  \half  \mu_0 \, {{\vep}_{zz}}^2 + \half \mu_1 
({{\vep}_{xx}}^2 +2{{\vep}_{xy}}^2 +{{\vep}_{yy}}^2)\nonumber 
\\ && + \half \left( \mu_2- \frac{D_2^2}{4 
D_1} \right) ({{\vep}_{xz}}^2 +{{\vep}_{yz}}^2) \; .
\label{renormpeterF2} 
\end{eqnarray}

The modulus $\mu_2$ is renormalised to  $\mu_2- {D_2^2}/{(4 D_1)}$ 
which by the GL theorem must be zero, thus establishing a relation 
between the constants   $\mu_2$, $D_1$, and  $D_2$.  The molecular 
model, required below for finite deformations, produces linear 
continuum limiting values which also give:
\begin{equation}
\mu_2^{\rm R} =\mu_2 - \frac{D_2^2}{4 D_1} \rightarrow 0 \; .
\label{mu2r} 
\end{equation}

 Olmsted \cite{olmsted} first 
proposed this continuum mechanism behind the Golubovic--Lubensky 
theorem: shape depends on the orientation of an internal (nematic) 
degree of freedom, the rotation of which causes a natural shape 
change at zero cost for suitable solids. A general discussion of 
the GL argument and its extension to semi-softness and thresholds 
to rotation is given in \cite{JMPS}.  The experimental evidence 
for mechanically soft distortions is also discussed there. 
 
One can picture the continuous rotation generating mechanical 
distortion at low energy cost, see Fig.~\ref{deG2}.  The natural 
long axis of the body, as defined by the principal axis of 
molecular shape, rotates in an attempt to follow the apparent 
extension axis. To the extent that macroscopic shape change can 
thus be imitated, there is no real accompanying distortion of 
polymer shape. We quantify this picture below when considering 
non-linear elasticity theory. 
 
\section{Finite soft elasticity}
 A simple extension of classical rubber elasticity theory 
to nematic elastomers gives for the free energy density:
\bea 
F=\Free \label{trace}
\eea
where $\mu$ is the rubber elastic shear 
modulus in the isotropic phase and
$\lambda_{ij} = \partial R_i/\partial R_{0j} $
is the homogeneous deformation (gradient).
If we take $\matr{\lambda}$ to be $\matr{\lambda}=\matr{\delta}+\matr{u}$, then
the symmetric part of $\matr{u}$ is identical to $\matr{\tilde{\varepsilon}}$ of
(\ref{peterF}) in the infinitesimal limit.

The chains are no longer 
characterised by spherical Gaussian distributions as in the 
classical case, but in general by anisotropic distributions. Thus 
the effective step length tensors  $\llm\s{0}$ initially and 
$\llm$ currently after a distortion $\lm$ are prolate (or oblate) 
spheroids defining the second moments that characterise the 
Gaussian distribution of chain spans $\vec{R}$ in the the network, 
that is $\left< R_i R_j \right> = \frac{1}{3} \ell_{ij} L$ where 
$L$ is the arc length of a chain. Thus a measure of the mean size of a 
chain is $\llm^{1/2}$ where we shall soon define the roots of 
tensors more carefully.  The tensor $\llm$ has one principal value 
$\lz$ along $\vec{n}$ and $\lp$ perpendicular to $\vec{n}$; thus 
$$
    \llm = \left(\begin{smallmatrix} \lp & 0 &0 \\ 0 & \lp &0 \\ 0 & 
                0& \lz \end{smallmatrix} \right)
          =\lp\left(\begin{smallmatrix}1& 0 &0 \\ 0 & 1 &0 \\ 0 & 0& r 
                \end{smallmatrix} \right),\nonumber
$$
where $r = \lz/\lp$.  The extracted $\lp$ 
factor from $\llm_0$ cancels with the  $1/\lp$ factor 
extracted from $\llm^{-1}$ when both tensors appear together in 
the Trace formula (\ref{trace}) and we can henceforth just consider the $\llm$ 
tensors in their reduced form that simply depends on the intrinsic 
anisotropy $r$. Anisotropy varies between $r=1.1 \rightarrow 60$. 
In this model of rubber elasticity the spontaneous elongation, 
$\lambda\s{m}$,  on going from the isotropic to the nematic phase 
turns out to be $\lambda\s{m} = r\sp{1/3}$ and is thus a direct 
measure of $r$.  Indeed spontaneous elongations in the range of 
$\lambda\s{m} \sim 1.03 - 4.00$ are observed. 

 Now distortions 
$\lm$ are no longer small, but must continue to respect volume 
conservation, $\rm{Det}(\lm) = 1$ in the non-linear regime.  The 
directors $\vec{n}\s{0}$ and $\vec{n}$, of the initial and current 
nematic states, may be greatly rotated from each other. 

The remainder of this paper is concerned with explaining the 
character of the soft modes within finite  elasticity theory and 
extending this picture to that of soft modes in biaxial nematic 
elastomers.  We explore the role of rotations and the connections 
between the isolation of rotational components of strain at finite 
amplitude (the spherical decomposition theorem) and the relation 
of the roots of tensors to this picture.

\section{Uniaxial softness}\label{uniaxial}
Consider the strain \cite{olmsted}: 
 \bea \lm =
\llm^{1/2} \cdot \matr{W}_{\alpha} \cdot \llm^{-1/2}\s{0} \; , 
\label{softdeform} 
 \eea
where $\matr{W}_{\alpha}$ is an arbitrary rotation by an angle 
$\alpha$. There are two continuous degrees of freedom describing the rotation
connecting $\n$ and $\n_{0}$ and three degrees of freedom for the
rotation $\matr{W}_{\alpha}$. Hence, the strain $\matr{\lambda}$ is
described by five continuous degrees of freedom and thus represents a large
set of deformations.

If we insert such a strain into the Trace 
formula (\ref{trace}), as well as its transpose $\lm^{\sf T}$ 
(equivalent to $\llm\s{0}^{-1/2}\cdot \matr{W}_{\alpha}^{\sf T} 
\cdot \llm^{1/2}$ since the $\llm$ are symmetric) we obtain: 
 \bea F\s{el} &=&
\half \mu \Tr{\llm\s{0} \cdot \llm\s{0}^{-1/2} \cdot \matr{W}^{\sf 
T}_{\alpha} \cdot \llm^{1/2} \cdot \llm^{-1} \cdot \llm^{1/2} 
\cdot \matr{W}_{\alpha} \cdot \llm^{-1/2}\s{0} }\nonumber\\ 
&\equiv& \half \mu \Tr{\deltam} = \frac{3}{2} \mu \; . 
\label{Olmsoft} 
 \eea
Cancelling the middle section, $\llm^{1/2} \cdot \llm^{-1} \cdot 
\llm^{1/2} = \deltam$, allows the $\matr{W}$-terms to meet and 
gives $\matr{W}^{\sf T}\cdot \matr{W} = \deltam$.  Likewise 
disposing of the $\llm\s{0}$ terms, one obtains the final 
expression $F\s{el} = \frac{3}{2}\mu$. This is identical to the 
free energy of an undistorted network. The non-trivial set of 
distortions $\lm$ of this particular form, equation (\ref{softdeform}), 
have not raised the energy of nematic elastomer. From the results 
$\rm{Det}(\matr{A}\cdot\matr{B})=\rm{Det}(\matr{A})\rm{Det}(\matr{B})$, 
$\rm{Det}(\matr{A}^{-1})=(\rm{Det}(\matr{A}))^{-1}$ 
and  $\rm{Det}(\matr{W})=1$ for all rotations $\matr{W}$, one can 
see that  $\rm{Det}(\lm)=\rm{Det}(\llm^{1/2} \cdot \matr{W}_{\alpha} 
\cdot \llm^{-1/2}\s{0} )=1 $, that is these soft modes are 
volume-preserving. 

We saw pictorially  in fig.~\ref{deG2} that, on applying a stretch 
perpendicular to the initial director, rotation of the chain 
distribution is accommodated by the very elongation we have 
applied, together with a shear. Two remarkable consequences 
immediately follow from nematic elastomer response via rotation: 

Impose an $x$-extension 
$\lambda_{xx}$.  All the accompanying distortions  must be in the 
plane of rotation, that is a transverse contraction 
$\lambda_{zz}$ and shears  $\lambda_{xz}$ and $\lambda_{zx}$ 
accommodate the rotation of the distribution.  No shears 
perpendicular to this plane, that is involving $y$-direction 
($\lambda_{yy}, \lambda_{yx}, \dots$ etc.) are needed.  For a 
classical isotropic elastomer $\lambda_{yy} = \lambda_{zz} = 
1/\sqrt{\lambda_{xx}}$ is demanded by incompressibility, whereas 
in soft elasticity there is no shrinkage in the $y$ direction 
($\lambda_{yy} = 1$) and the appropriate Poisson ratio is zero. 

\begin{figure}
\resizebox{0.45\textwidth}{!}{\includegraphics{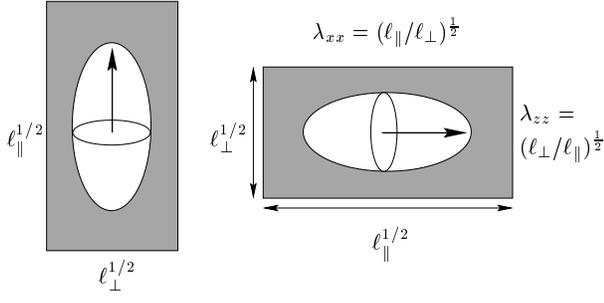}} 
\caption[]{Chain shape distribution before and after rotations. 
The extent of soft extensions ($\lambda_{xx}$) perpendicular to 
the initial director is set by the anisotropy of the molecular 
shape distribution.  The macroscopic dimensions are shown changing 
affinely with the distribution, for instance the $x$ dimension 
changing from $\sqrt{\lp}$ to $\sqrt{\lz}$. } \label{visual} 
\end{figure}

Fig.~\ref{visual} shows the initial and final states of which the 
shear and partial rotation of fig.~\ref{deG2} is an intermediate 
step.
 Softness must come to an end when the rotation is complete 
and the $z$ dimension has diminished in the proportion 
$\lambda_{zz} = \sqrt{\lp/\lz}$ and the $x$ dimension extended in 
the proportion $\lambda_{xx} = \sqrt{\lz/\lp}$. The original 
sizes $\sqrt{\lz}$ and $\sqrt{\lp}$ have transformed to 
$\sqrt{\lp}$ and $\sqrt{\lz}$ respectively.  Thus softness would 
cease and director rotation be complete at $\lambda_{xx} = 
r^{1/2} \equiv \lambdam^{3/2}$. The strain $\lambdam = 
(\lz/\lp)^{1/3}$ is the spontaneous extension suffered on cooling 
to the nematic phase. Likewise one can imagine from an oblique 
form of figure \ref{visual} that if the initial director $\n_{0}$ 
(long axis of the shape ellipsoid) is not at $90^{\rm o}$ to the 
imposed strain, then rotation and softness is complete at a 
smaller $\lambda_{xx} < \lambdam$.

We give as a concrete example the set of soft distortions  
$\lm\s{soft} = \llm^{1/2}_{\theta}\cdot \llm\s{0}^{-1/2}$, simplified 
by the absence of the arbitrary rotation matrix 
$\matr{W}_{\alpha}$. 
 They
are simply characterised (parametrically) by the angle $\theta$ by 
which $\llm\s{0}$ is rotated to $\llm_{\theta}$, that is by which 
$\n\s{0}$ is rotated to $\n$.  Putting in the dyadic forms for 
$\llm\s{0}^{-1/2}$ and $\llm^{1/2}_{\theta}$ into $\lm\s{soft}$ 
gives: 
\bea
    \lm\s{soft} &=& (\deltam +(\sqrt{r}-1)\n\n^{\sf T})\cdot
        (\deltam + (\frac{1}{\sqrt{r}} - 1)\n\s{0}\n_{0}^{\sf T})\nonumber\\
    &=& \deltam + (1/\sqrt{r} - 1)\n\s{0}\n_{0}^{\sf T} +
        (\sqrt{r}-1)\n\n^{\sf T} \nonumber\\
    && +(\n.\n_{0})(2 - \sqrt{r} - 1/\sqrt{r} )\n\n_{0}^{\sf T} 
        \; . \label{parasoft}
\eea
If $\n\s{0}$ is along $\z$ and is rotated by $\theta$ toward $\x$, 
it becomes $\n = \z \cos\theta + \x\sin\theta$.  We can write down 
a particular representation of $\lm\s{soft}$ (using the notation $s 
\equiv \sin\theta$ and $c \equiv \cos\theta$): 
\bea
    \lm\s{soft} &=& \z\z^{\sf T}(1 - (1 - \frac{1}{\sqrt{r}})s^2) +
    \x\x^{\sf T} (1 + (\sqrt{r}-1)s^2)  \nonumber\\
    && + \y\y^{\sf T} + \x\z^{\sf T}(1-\frac{1}{\sqrt{r}}) s c +
        \z\x^{\sf T} (\sqrt{r} -1) s c \nonumber\\
    &\equiv&
        \begin{pmatrix}1 + (\sqrt{r}-1)s^2 & 0 &  (1- 1/\sqrt{r})s c\cr
                0 & 1 & 0   \cr
                (\sqrt{r} -1) s c & 0 & 1 - (1 - 1/\sqrt{r})s^2
        \end{pmatrix}
    \; .\label{parasoft2}
 \eea
The soft modes are neither  simple nor  pure shear, but a mixture 
of the two.  
Accordingly, the soft modes have an element of body rotation in addition to
elongations, compressions and pure shears. The importance of body rotations is
discussed before and after (\ref{Oeq}). The degree of body rotation is given in
section \ref{rss}.
Note that the extensional and compressional strains 
$(\lambda_{xx} - 1)$ and $(1 -\lambda_{zz})$  are both 
proportional to $\sin^2 \theta$. Thus the infinitesimal strains 
$u_{zz}$ and $u_{xx}$, at small rotations 
$\theta$, are proportional to $\theta^2$.  By contrast 
$\lambda_{xz}$ and  $\lambda_{zx}$ are proportional to $\sin 
\theta \, \cos \theta$ and hence the infinitesimals 
$u_{xz}$ and $u_{zx}$ are proportional to $\theta$ 
-- a lower order than $u_{xx}$ and $u_{zz}$. There 
is no relaxation along $y$, perpendicular to the plane of rotation 
$\vec{n}$, that is $\lambda_{yy} = 1$.  Also note that in the 
isotropic limit ($r=1$) this particular strain $\lm\s{soft}=\deltam $ 
while the general soft deformation (\ref{softdeform}) reduces to 
the null strain, $\matr{W}_{\alpha}$, a simple body rotation. 
Both are the trivial cases evidently preserving the elastic energy 
at its minimum. The soft modes become non-trivial deformations 
when the material becomes a nematic elastomer.

The soft modes start at no strain,  $\lm = \deltam$, and as the 
director $\theta$ rotates from 0 to $\pi/2$, they 
eventually end at
$$
\lm = \left(\begin{smallmatrix} \sqrt{r} & 0 &0 
\\ 0 & 1 &0 \\ 0 & 0& 1/\sqrt{r}  \end{smallmatrix} \right),
$$
that is an extension $\lambda_{xx} = 
\sqrt{r}$ and a transverse contraction $\lambda_{zz} = 
1/\sqrt{r}$. 

\begin{figure}
\centering 
\resizebox{.45\textwidth}{!}{\includegraphics{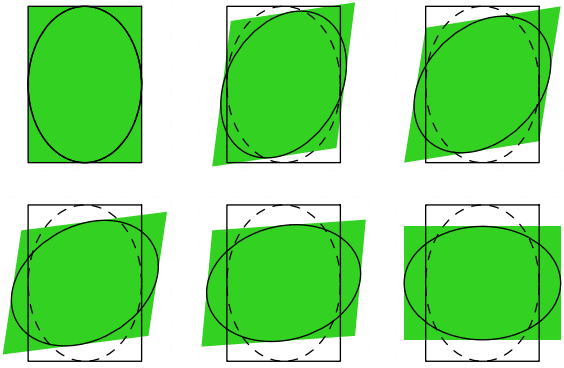}} 
\caption[]{Soft deformations of a nematic elastomer with 
anisotropy $r = 2$. The deformations correspond to director 
rotations of $\theta=0$, $\pi/6$, $\pi/4$, $\pi/3$, $5\pi/12$ and 
$\pi/2$ which parametrically generate the distortions as discussed 
above. Note that the prolate spheroid characterising the 
distribution of chains, embedded in the distorting solid (shaded 
and shown in section), when rotated by $\theta$ can be 
accommodated without distortion. The reference, undeformed body is 
shown in outline; the original distribution shown dashed.} 
\label{softstrains_b} 
\end{figure}

The director rotation is taken up by a shape change so that there 
is no entropically expensive deformation of the chain distribution 
as when a conventional elastomer deforms.  The shape tensor's 
anisotropy $r = \lz/\lp$ characterises the ratio of the mean 
square size along the director to that perpendicular to the 
director.  The square root of this ratio, $\sqrt{r}$, gives the 
characteristic ratio of average dimensions of chains in the 
network.  During a soft deformation, the solid must change shape 
such that the rotating ellipsoid $\llm^{1/2}$, characterising the 
physical dimensions of the distribution of chains, is accommodated 
without distortion. The ellipsoid is $\R\cdot\llm^{-1}\cdot \R = 
1$, or in a principal frame, $ x^2/r + z^2 = 1$, that is in 
section the ellipse has semi-major axes of $\sqrt{r}$ and $1$. 
Fig.~\ref{softstrains_b} illustrates soft deformations of a 
nematic elastomer with chains of anisotropy $r =2$.  The 
distortions are parameterised by the director rotation, $\theta$, 
which ranges between $0$ and $\pi/2$. The chain shape tensor, 
$\llm^{1/2}$, rotates without distortion, just fitting into the 
solid into which it embedded. 
The soft deformations of fig.~\ref{softstrains_b} and 
equations (\ref{parasoft}) and (\ref{parasoft2}) of the body are in 
general non-symmetric. Below we decompose these into pure shears 
plus rotations. One can think of $\lm\s{soft} = 
\llm^{1/2}_{\theta}\cdot \llm\s{0}^{-1/2}$ as converting the 
initial ellipsoid of fig.~\ref{softstrains_b} to a sphere by the 
action of the inverse   $ \llm\s{0}^{-1/2}$ and then recreating an 
ellipsoid at an angle $\theta$ with the action of  
$\llm^{1/2}_{\theta}$.  This is precisely the scheme of 
\cite{deSimone,Lubensky:01} who consider an isotropic reference 
state (the intermediate created after the action of  $ 
\llm\s{0}^{-1/2}$).  We discuss this rotational invariance again 
below.
 
\section{Biaxial Softness}\label{biaxialsoftness}

Biaxial nematic phases are rare. They have been found (Finkelmann 
{\it et al} \cite{Finkbiaxial,Finkbiaxial2}) in nematic polymers 
since one has more complex molecular structural possibilities.  In 
principle such polymers could be used to make biaxially nematic 
elastomers.  They would have rich mechanical properties. 

\begin{figure}[h!]
\centering 
\resizebox{.4\textwidth}{!}{\includegraphics{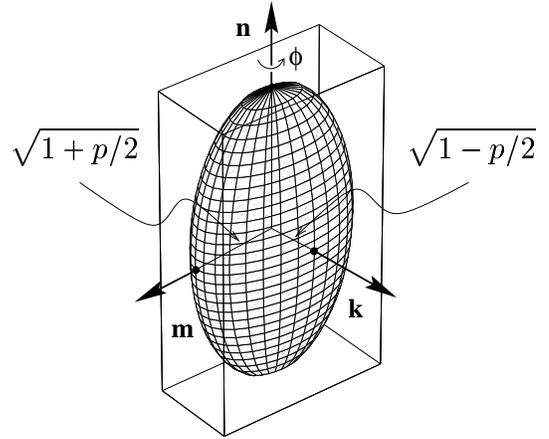}} 
\caption[]{The shape ellipsoid for a biaxially nematic polymer. 
The section perpendicular to $\n$ is not circular, but has 
semi-axes of length $\sqrt{1\pm p/2}$.  Rotations $\phi$ about 
$\n\equiv\n_0$ generate distortions in the $xy$ plane that softly 
accommodate the non-circular shape as it rotates. } 
\label{biaxialellipsoid} 
\end{figure}

The shape tensor of a biaxial polymer is
$$\llm = 
    \left(\begin{smallmatrix} \ell_1 & 0 &0 \\ 0 & \ell_2 &0 \\ 0 & 0& 
    \lz 
    \end{smallmatrix} \right).
$$
Now the step lengths in the two directions perpendicular
to $\n$, $\ell_1$ and $\ell_2$, are distinguished.  The tensor is
reduced by taking out a factor of the mean perpendicular step 
length, $\lp = \half (\ell_1 + \ell_2)$, to give:
\bea
    \llm &=& 
        \left(\begin{smallmatrix} 1 + p/2&0&0 \cr 0&1-p/2 &0 \cr 0 & 0& r 
        \end{smallmatrix}\right)\nonumber \\
    & \equiv & r \n\n^{\sf T} +(1 + \frac{p}{2})\m\m^{\sf T} + (1- \frac{p}{2})
        \k\k^{\sf T} \nonumber\\
    &\equiv& \deltam + (r-1) \n\n^{\sf T} + \frac{p}{2}\m\m^{\sf T} - 
        \frac{p}{2}\k\k^{\sf T}\nonumber
\eea
where the axes of the ellipsoid are 
$\n$, $\m$ and $\k$ (with the latter a third perpendicular axis 
given by $\k = \n\times \m$), see fig.~(\ref{biaxialellipsoid}). 
The biaxiality is $p = 2\frac{\ell_1 - \ell_2} {\ell_1 + \ell_2} = 
\frac{\ell_1 - \ell_2}{\lp}$. 

With fixed $\n$, in the biaxial case one can rotate the 
anisotropic transverse section $m k$ of the square root of the 
shape tensor, $\llm^{1/2}$, and accommodate it without  distortion 
by inducing shape changes in the $mk$ plane. These distortions are 
soft for exactly the same reasons as in the uniaxial case when 
$\n$ suffered rotations in the $nk$ or $nm$ planes, see 
figs.~\ref{softstrains_b}. Given that there are (now in the Lab. 
frame) $\lambda_{xy}$ soft modes as well as the $\lambda_{zx}$
and $\lambda_{zy}$ forms (which are now no longer 
equivalent modes), the order of softness has become very much 
greater. 

\subsection{Explicit examples of soft modes arising from biaxiality} 

The situation is exactly parallel to that of the soft 
shears (\ref{parasoft}) and (\ref{parasoft2}). Now the soft modes 
are given by $\llm^{1/2}_{\phi}\llm^{-1/2}\s{0}$ where $\phi$ is 
instead the angle of rotation about the principal director $\n$. 
The tensor $\llm\s{0}^{1/2}$ is 
\begin{equation}
    \llm^{1/2}_0= \left(\begin{smallmatrix} 
    \sqrt{1 + p/2} & 0 &0 \\ 0 & \sqrt{1 - p/2} &0 \\ 0 & 0& \sqrt{r} 
    \end{smallmatrix} \right),\label{rootbiax}
\end{equation}
which will then be rotated by $\phi$. By analogy with 
fig.~\ref{softstrains_b}, the anisotropy in the $x y$ part of the 
matrix which is to be rotated is $\sqrt{r_{\bot}} = \sqrt{\frac{1 
+ p/2}{1-p/2} } \equiv \sqrt{\frac{\ell_1}{\ell_2}}$. This 
measures the anisotropy of actual (root mean square) chain 
dimensions in the plane perpendicular to the director.  The 
resulting soft shears in the $xy$-plane arising from rotations $\phi$ about
${\bf z}={\bf n}_0$ are: 
\bea
    \lm\s{biaxial} =
    \left(\begin{smallmatrix}
    1-(1-1/\sqrt{r_{\bot}})s^2_{\phi}
    &(1-\sqrt{r_{\bot}})s_{\phi} c_{\phi}&0\cr
    (1/\sqrt{r_{\bot}}-1)s_{\phi} c_{\phi}
    & 1+(\sqrt{r_{\bot}}-1)s_{\phi}^2 &0\cr
    0 &0 &1 \end{smallmatrix}\right) \; . \label{unibiax}
\eea
More complicated soft shears are possible if these shears are 
combined with those previously found in the uniaxial case.  For 
example, in the uniaxial case,  section (\ref{uniaxial}) above, 
subsequent rotation about $\n$ after the rotation of $\n$ itself 
had no effect since there was symmetry about $\n$.  Now that this 
is lost, such rotations will reorient the perpendicular directors 
$\m$ and $\k$ and thereby induce additional soft shears. 

\subsection{General biaxial softness}

\begin{figure}[h] \centering
\resizebox{0.4\textwidth}{!}{\includegraphics{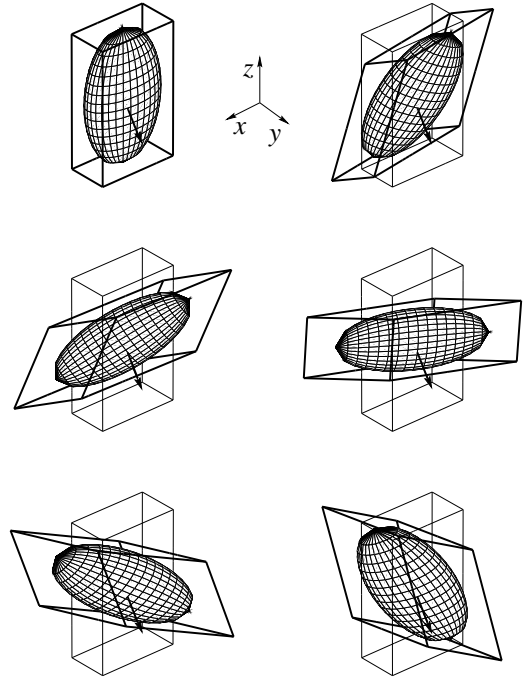}} 
\caption[]{Rotation of the chain distribution around an axis 
(arrow) in the $xy$-plane at an angle of $\pi/3$ with the 
$x$-axis. The figure shows intermediate states of rotation at an 
angle of 0, $\pi$, 
$\frac{1}{6}\pi$,$\frac{1}{3}\pi$,$\frac{1}{2}\pi$,$\frac{2}{3}\pi$ 
and $\frac{5}{6}\pi$. Fig. (\ref{sequenceend}) shows the final 
stage of a rotation by an angle of $\pi$. The fine outline shows 
the initial body shape, whereas the highlighted outline 
illustrates the body deformation as the the ellipsoid of the 
internal chain distribution rotates. Here, we have chosen $p=6/5$ 
and $r=200/45$, or, equivalently,  the principal axes of the 
ellipsoid to be 6, 3 and 10 respectively.} \label{sequence} 
\end{figure}

\begin{figure}[h] \centering
\resizebox{0.4\textwidth}{!}{\includegraphics{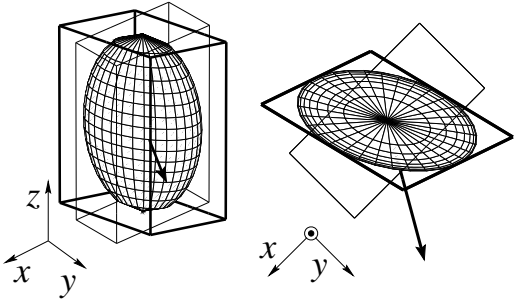}} 
\caption[]{Rotation of the chain distribution by $\pi$ around an 
axis in the $xy$-plane at an angle of $\pi/3$ with the $x$-axis. 
The figure shows two views of the same situation: the 
final ellipsoid together with the outlines of the bodies which 
indicate the corresponding soft mode deformation. The outline of 
the  deformed body is highlighted. The parameters are the same as in fig.
(\ref{sequence}).}
\label{sequenceend} 
\end{figure}

If, however, the axis of rotation does not coincide with a 
principal axis as in the previous examples, we discover soft modes 
of lower symmetry, which only lead back to an undeformed body 
after the chain distribution has been rotated by $2\pi$. Consider 
as an example of reduced symmetry the situation when the chain 
distribution is rotated around a general axis in the $xy$ plane at 
an angle $\beta$ to the $x$ axis (fig. \ref{sequence}): a rotation 
of $\pi$ (fig. \ref{sequenceend}) transfers the director ${\bf n}$
(along the $z$-axis) 
into $-{\bf n}$, but the unit vectors ${\bf m}$ and ${\bf k}$ 
along the other two principal axes end up in a general position in 
the $xy$-plane.  The same final position of the ellipsoid of chain 
distribution can be obtained by a rotation around the $z$-axis by 
an appropriate angle. The corresponding body deformation leaves 
all $z$-coordinates invariant, but transforms the the 
$xy$-coordinates as if they have suffered a body rotation of 
$2 \beta$ about $z$. 

If we were to choose the axis of rotation in an arbitrary way, we would not
observe this feature of invariant $z$-components after a rotation of $\pi$.
Only a full rotation of $2\pi$ would lead back
into a state of higher symmetry, in this case, of course, the identity.

By rotating the ellipsoid of chain distribution, we create soft 
modes, which are closely related to the underlying symmetry, which 
is the crystal class of the biaxial ellipsoid, the dipyramidal 
orthorhombic class. 

Comparing with equation (\ref{softdeform}), we see that so far we have set the
rotation $\matr{W}_{\alpha}$ to be the identity. If, however, we were to allow a
general rotation, we would find an even larger set of deformations, in fact
described by six degrees of freedom: the additional rotations
$\matr{W}_{\alpha}$ and the rotations connecting the principal frames of
the initial and final ellipsoids each introduce three degrees of freedom.

\section{Rotations, symmetric strains and the roots of tensors}\label{rss}

We have seen in equations (\ref{peterF})--(\ref{renormpeterF2}) yielding
softness that rotations are vital to understanding soft ela\-sti\-ci\-ty. The
cartoon of fig. 5 shows that soft shears are not symmetric and hence there must
be a component of body rotation present. Here we explore the role of rotations
in symmetry requirements. We then explicitly extract the rotational component
from general distortions and from soft deformations in particular. The
requirements for extracting the rotational components of the strain tensor are
intimately related to finding the square roots of tensors which we discuss here
since they are used in constructing soft deformations.

Any general matrix $\matr{M}$ can be broken down
into pro\-ducts $\matr{S}\cdot \matr{O}$
or $\matr{O}\cdot \matr{S}'$ of symmetric matrices
$\matr{S}$ and $\matr{S}'$ and orthogonal matrices
$\matr{O}$ and $\matr{O}'$,
which can be restricted to proper rotations.
In effect, one has a pure shear preceded or followed by a body rotation (the 
Polar Decomposition Theorem \cite{HornJohnson}).

Note that in the infinitesimal case, we have
$\matr{S}=\matr{1}+\matr{\delta S}$ and $\matr{O}=\matr{1}+\matr{\delta O}$,
hence $\matr{\delta S}=\matr{\delta S}^{\sf T}$ and
$\matr{\delta O}=-\matr{\delta O}^{\sf T}$. Analogous statements hold for the
matrices $\matr{S}'$ and $\matr{O}'$.
This shows that the rotation matrices
$\matr{O}$ and $\matr{O}'$ are related to the antisymmetric part of the
deformation in the infinitesimal limit.

Applied to the deformation
$\lm$, we write:
\bea
    \lm = \lm\sp{L}\cdot  \matr{V} \;\;\;\ {\rm or} \;\;\; 
            \matr{U} \cdot \lm\sp{R},
\eea
where the rotations are denoted by \matr{V} or \matr{U} depending
upon whether  they act 
on the reference or target spaces respectively of the deformation.  
The form of the accompanying symmetric deformations 
will depend on the order; they are denoted by 
$\lm\sp{R}$ and $\lm\sp{L}$ respectively and yield the 
Cauchy-Green tensors \matr{C} and \matr{B} respectively:
$\matr{C}=\lm^{\sf T}\cdot\lm$ and $\matr{B}=\lm\cdot\lm^{\sf T}$. 

We summarise classical elasticity theory to highlight the 
differences with nematic rubber elasticity.
 The deformation (gradient) is: 
 \bea
 \lambda_{ij} = \partial R_i/\partial R_{0j} \; .
 \eea
If the target space ($S_T$) deforms under rotations represented by 
the matrix \matr{U}, as $\R' = \matr{U}\cdot \R$, and the 
reference space  ($S_R$) deforms under rotations $\matr{V}$ as 
$\R_{0}' = \matr{V} \cdot \R_{0}$, then the deformation  tensor deforms 
as 
\bea
    \lambda'_{ij} &=& U_{ik}\partial R_k / \partial R_{0l} 
    V\sp{\sf T}_{lj}\label{transform} \\
    \matr{\lambda}' &=& 
    \matr{U}\cdot \matr{\lambda} \cdot 
    \matr{V}\sp{\sf T}\label{transform1}\quad {\rm or}\quad  
    \matr{\lambda}
    = \matr{U}\sp{\sf T}\cdot \matr{\lambda}' \cdot 
    \matr{V}\label{transform2}\; . 
\eea 
Thus $\matr{\lambda}$ records the character of both the target and 
reference states, a property that will be essential in non-ideal 
nematic elastomers where an isotropic reference state cannot be 
reached. The connection with both spaces is quite different in 
character from that of the Cauchy tensors. Thus the combination
\bea
    \matr{C} &=& \matr{\lambda}\sp{\sf T} \cdot 
    \matr{\lambda} = \matr{V}\sp{\sf T} \cdot {\matr{\lambda}'}^{\sf T} \cdot 
    \matr{U}\cdot \matr{U}\sp{\sf T} \cdot \matr{\lambda}' \cdot \matr{V}
        \nonumber\\
    &=& \matr{V}\sp{\sf T} \cdot {\matr{\lambda}'}^{\sf{T}} \cdot \matr{\lambda}' 
    \cdot \matr{V} = \matr{V}\sp{\sf T} \cdot \matr{C}' \cdot \matr{V} 
\eea
is manifestly invariant under body rotations \matr{U} of the 
final (target) space $S_T$ and transforms as a second rank tensor 
in $S_R$. Since isotropic systems are invariant to rotations 
\matr{V} of $S_R$, the system's final energy must be invariant to 
rotations of $S_R$. Thus $F$ is a function of the rotational (in 
$S_R$) invariants of $\matr{C}$ and is assured by the above of 
being invariant under rotations of $S_T$. As an example, the 
isotropic rubber elastic free energy, setting $\llm\sp{0} = 
\llm\sp{-1}= \deltam$ in equation (\ref{trace}) is 
 \bea F &=& \half \mu
\Tr{\matr{C}} =  \half \mu 
\Tr{\matr{V}\sp{\sf T}\cdot\matr{C}'\cdot\matr{V} } \cr &=& \half \mu 
\Tr{\matr{C}'\cdot\matr{V}\cdot\matr{V}\sp{\sf T} } =  \half \mu 
\Tr{\matr{C}'} \eea (by cyclical properties of the trace). 
Likewise, $\matr{B} = \matr{\lambda}\cdot\matr{\lambda}\sp{\sf T}$, is 
invariant to rotations of the reference state and transforms like 
a second rank tensor in the target state. 

Thus we see that ``objectivity", frame indifference, is built into 
classical elasticity theory from the outset.  Nematic elastomers 
are much more subtle.  In continu\-um theory we have seen that 
rotations and symmetric shears enter, both separately (through
the de Gennes $D_1$ term) and coupled 
(through the $D_2$ term).  In finite elasticity, we see 
in the Trace formula that the initial and final orientations of 
the solid and its directors enters via the tensors $\llm\sp{0}$ 
and $\llm\sp{-1}$ and there are not combinations like $\matr{B}$ 
and $\matr{C}$ which eliminate rotations.  For instance, inserting  
$\matr{U} \cdot \lm\sp{R}$ into the trace result, one obtains: 
\bea
    F = \half \mu \Tr{\llm\s{0}\cdot \lm^{\rm R\;{\sf T}}\cdot [
    \matr{U}\sp{\sf T} \cdot  \llm^{-1}\cdot \matr{U} ] \cdot \lm\sp{R} } 
\eea
where the $[\dots ]$ have been inserted to emphasise that a 
body rotation of $S_T$ effectively adds to the rotation of $\n$.  
That is, a new $\llm '$ evolves:  ${\llm'}^{-1} = \matr{U}^{\sf T} 
\llm^{-1} \matr{U}$.  (The additional rotation is not necessarily 
coaxial with that which took $\n\s{0}$ to $\n$.) Not unexpectedly, 
we see the effect of $\matr{U}$ compounded with the rotations 
implicit in $\llm^{-1}$ since they  both live in the target space. 

Other approaches have been taken \cite{deSimone,Lubensky:01} to 
soft elasticity which apparently circumvent the necessity to 
follow orientations in both spaces, and restore objectivity.  One 
can measure all deformations $\lm$ from an isotropic reference 
state, that is there is encoded into the $\lm$ first a spontaneous 
deformation to current conditions of  temperature, then a 
deformation imposed with respect to this intermediate state.  
Under these conditions the free energy must automatically be 
invariant under  operations of $\matr{V}$, since the reference 
state is isotropic and without a director $\vec{n}\s{0}$ to keep 
track of.  Difficulties arise then when nematic elastomers are 
only {\em semi-soft}, that is, they do not deform entirely at 
constant free energy, because they do not have a high temperature 
isotropic reference state.  However they do suffer director 
rotation, very low energy trajectories in $\lm$ space and a lack 
of $\lambda_{yy}$ relaxation perpendicular to the plane of 
$\vec{n}$'s rotation \cite{semi_soft}.  In these cases, the 
complete cancellation  rendering $\mu_2^{\rm R} =\mu_2 - 
\frac{D_2^2}{4 D_1} \rightarrow 0$ fails, but nevertheless 
deformations are qualitatively soft and one has to keep track of 
both directions  $\vec{n}\s{0}$ and $\vec{n}$ as strain evolves. 

\subsection{Square roots of tensors and the Polar Decomposition Theorem}
 We quote the classical conditions\cite{HornJohnson} for the square roots 
of tensors, that arise too in the all-important condition for 
polar decomposition that is required for nematic elastomers. 

If $\matr{A}$ is non-singular, with $\mu$ distinct eigenvalues 
and with $\nu$ Jordan blocks, then  $\matr{A}$ has $\ge 2^{\mu}$ 
and $\le 2^{\nu}$ non-singular square roots.  At least one of 
these roots is a polynomial in $\matr{A}$.  

The proof of the Polar Decomposition Theorem (PDT)
also offers a practical algorithm for decomposition: take 
a non-singular $\matr{\lambda}$ and construct the manifestly 
symmetric $\matr{\lambda}\sp{\sf T}\matr{\lambda}$. Take the square 
root, $\matr{G}$, of $\matr{\lambda}\sp{\sf T}\matr{\lambda}$ that is 
a polynomial in $\matr{\lambda}\sp{\sf T}\matr{\lambda}$.  Since 
$\matr{\lambda}\sp{\sf T}\matr{\lambda}$  is symmetric, then so is  
$\matr{G} = \sqrt{\matr{\lambda}\sp{\sf T}\matr{\lambda}}$.  Then 
define $ \matr{Q} = \matr{G}\sp{-1}\matr{\lambda}$.  Clearly   $ 
\matr{Q}\sp{\sf T} \matr{Q} = \deltam$, that is $\matr{Q}$ represents 
rotations.  From this one recovers  $ \matr{\lambda}= 
\matr{G}\;\matr{Q} $. 
 
\subsection{Symmetric strain and body rotation of soft modes}\label{softdecomp}

We investigate the special case of uniaxial soft modes in the $xz$ 
plane. They are achieved by rotating the chain distribution around 
the the $y$ axis. The resulting deformation keeps the 
$y$-components of the body constant. Hence these soft modes are 
effectively $2\times 2$ and can be easily related to equation 
(\ref{parasoft2}) and fig. \ref{softstrains_b}. 

An explicit example of the 
PDT for decomposing a general $xz$-distortion $\lm$ into a 
combination of symmetric distortion, $\lm\sp{R}$, followed by a body 
rotation, $\matr{U}$, about the $y$-axis by an angle $\alpha$ is:
 \bea
    \lm_{\rm soft}=
    \begin{pmatrix} \lambda_{xx}&\delta\cr\delta '&\lambda_{zz} \end{pmatrix}
    = \begin{pmatrix} c&s\cr -s& c \end{pmatrix} \, \begin{pmatrix}  
    a&d \cr d &b \end{pmatrix} \equiv \matr{U}\cdot \lm\sp{R} \; .  \nonumber
\eea
(with $s= \sin\alpha$ and $c= \cos\alpha$). One can 
confirm that the rotation is:
\bea \tan\alpha = (\delta - \delta 
    ')/(\lambda_{xx} + \lambda_{zz}) \label{tantheta}
\eea and thus for instance for $\sin\alpha$: \bea 
    \sin\alpha = \frac{1}{\Delta}(\delta - \delta')
    \ {\rm with} \
    \Delta^2=(\lambda_{xx} + \lambda_{zz})^2 + (\delta -\delta')^2\;
        .\nonumber 
 \eea
Of course there is no body rotation for 
$\lm$ symmetric, that is $\delta = \delta '$. One can also confirm 
that the symmetric shear tensor is :
\bea
    \lm\sp{R} = \frac{1}{\Delta}
        \left(\begin{smallmatrix}
        \lambda_{xx}(\lambda_{zz} + \lambda_{xx})-
            \delta'(\delta - \delta ')
        &\lambda_{xx}\delta +\delta'\lambda_{zz}
        \cr
        \lambda_{xx}\delta+\delta'\lambda_{zz} 
        &\lambda_{zz}(\lambda_{zz} + \lambda_{xx})
            +\delta (\delta - \delta ')
        \end{smallmatrix}\right).\qquad 
    \label{symdecomp}
\eea 

The results (\ref{tantheta}) and 
(\ref{symdecomp})  break the soft modes down into a symmetric 
shear $\lm\sp{R}$ followed by a body rotation $\matr{U}$ 
through an angle $\alpha$ about the $y$-axis. Thus the soft mode 
is $\lm\s{soft} =\matr{U}\cdot\lm\sp{R}$. We continue to 
parameterise them with the director rotation $\theta$. The body 
rotation is through an angle $\alpha$ given by 
\bea
    \tan\alpha &=& 
    \frac{\sin \theta \, \cos \theta (\sqrt{r} -1)^2}
   {2\sqrt{r} + \sin^2 \theta (\sqrt{r} -1)^2}\nonumber\\
  &=& \frac{\tan\theta(\sqrt{r} -1)^2}{2\sqrt{r} + \tan^2\theta(r+1)}
  \label{rotalpha}
 \; .
\eea

For small shear (small director rotation $\theta$), the component 
of body rotation is small too and proportional to $\theta$: 
\bea
    \alpha \approx \theta \frac{(\sqrt{r} -1)^2}{2\sqrt{r}}
    \; .\label{alphasmalltheta}
\eea
For large rotations, $\theta \approx \pi/2$, the rotation
$\alpha$ vanishes as we have seen in fig. \ref{softstrains_b}. 
Thus it is only at first that body rotation plays a part in 
accommodating the rotating chains. As the rotation of chains approaches
$\pi/2$, the body is simply extended or compressed along the principal axes of
the original chain distribution.

The corresponding symmetric shear strain $d$, the off-diagonal 
component of $\lm^{R}$, is 
\bea
    d = \frac{1}{\Delta}\frac{\sin\theta\,\cos\theta (r-1)}{\sqrt{r}}\nonumber 
\eea
which, for small distortions is also proportional to $\theta$, bearing in mind
that $\Delta$ is constant at first order in $\theta$: 
\bea
d = \frac{1}{\Delta}\frac{\theta(r-1)}{\sqrt{r}}.
\label{dsmalltheta}
\eea

\begin{figure}[h] \centering
\resizebox{0.45\textwidth}{!}{\includegraphics{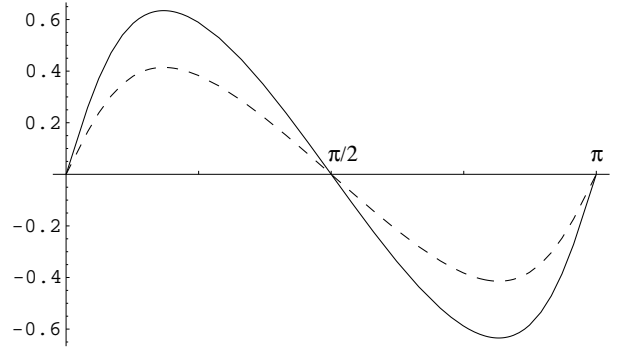}} 
\caption[]{Off-diagonal element $d$ of the symmetric shear 
$\matr{\lambda}\sp{R}$,  and angle $\alpha$ of the rotation  
$\matr{U}$ (dotted line) plotted against director rotation 
$\theta$ for the soft deformations of fig.~\ref{softstrains_b} for 
a nematic elastomer with anisotropy $r = 20$.} \label{symrot} 
\end{figure}

As we have seen in fig. \ref{softstrains_b}, it also vanishes at 
$\theta = \pi/2$, where no further accommodation of the shape 
tensor by rotation is possible. In fig. \ref{symrot}, we show the 
rotation and  off-diagonal element of the symmetric shear 
occurring during soft deformations between $\theta = 0$ and $\pi$. 
Notice that both quantities are 
linear in $\theta$ for small distortions. As the director of the 
chain distributions is rotated beyond $\pi/2$, the solid body 
rotation $\alpha$  and the off-diagonal element of the shear $d$ 
become negative. At a director rotation of $\pi$, the original 
body shape is recovered, as it should be in an uniaxial nematic 
system under reflection of the director ${\bf n}\rightarrow -{\bf 
n}$. 

\begin{figure}[h] \centering
\resizebox{0.4\textwidth}{!}{\includegraphics{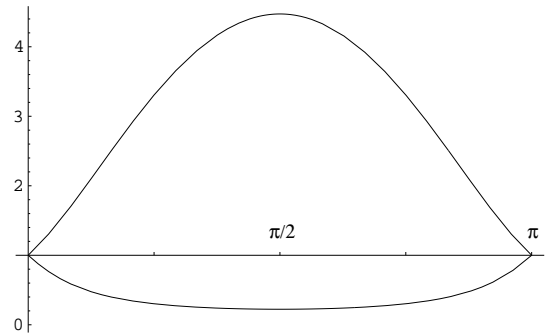}} 
\caption[]{The two eigenvalues of the pure shear 
$\matr{\lambda}\sp{R}$ for a nematic elastomer with anisotropy 
$r=20$. For a director rotation of $\pi/2$, the soft mode 
deformation becomes a pure shear with extension and compression 
ratio $\sqrt{r}$ or $1/\sqrt{r}$ respectively.} \label{eigen} 
\end{figure}

Similarly, the behaviour of the pure shear is instructive: for 
this purpose, we plot in fig. \ref{eigen} the two eigenvalues of 
$\matr{\lambda}^{\rm R}$, which serve as a good measure of the net 
presence of shear. 
  Note that, due to incompressibility, the
product of the two eigenvalues is constant to 1. Both eigenvalues are linear for
small $\theta$, saturate at values $\sqrt{r}$ and $1/\sqrt{r}$ respectively for
a director rotation of $\pi/2$, where the soft mode
becomes a simple compression and extension along the principal axes.
At a rotation of $\pi$, we
recover the eigenvalues of the identity matrix.

>From equation (\ref{parasoft2}), we see that the compression and 
extension in the $x$ and $y$ directions are quadratic in $\theta$ 
for small director rotations. In other words, $u_{zz}$ and 
$u_{xx}$ too are proportional to $\theta^2 \propto \alpha^2 
\propto d^2$ (by equations (\ref{dsmalltheta}) and 
(\ref{alphasmalltheta})). 

\section{conclusions}
Soft deformations of nematic elastomers result from the rotation 
of the anisotropic chain shape distribution without distortion and therefore
without rubber elastic free energy cost.
This is by contrast to classical rubber, where distorsion of the distribution
lowers entropy and raises the free energy.  
Explicit forms of these soft modes are derived in the case of 
biaxial nematic elastomers using the Olmsted method of the square 
roots of tensors developed for the uniaxial case.  For both cases 
we show geometrically what these deformations look like.  They 
correspond to rotations of prolate distributions being 
accommodated by elastic strains of the body they are inscribed 
into.  

The elasticity of nematic elastomers depends, unlike in classical 
elastomers, on body rotations (since these can be with respect to 
the underlying nematic director). At finite strains it is 
important to isolate the rotational component of the generally 
non-symmetric deformations.  We make contact with the Polar 
Decomposition Theorem in this context.  In doing so we discuss the 
roots of tensors that are employed in finding the manifold of soft 
deformation tensors. 

\begin{acknowledgments}
S. K. acknowledges the support of an Overseas Research 
Scholarship, of the Cavendish Laboratory and of Corpus Christi 
College. M. W. thanks the A. v. Humboldt Foundation for the award 
of a Humboldt Research Prize. 
\end{acknowledgments}

\bibliographystyle{prsty}


\end{document}